\documentclass{article}
\pdfoutput=1

\usepackage{ijcai21}

\usepackage{times}
\usepackage{soul}
\usepackage{url}
\usepackage[hidelinks]{hyperref}
\usepackage[utf8]{inputenc}
\usepackage[small]{caption}
\usepackage{graphicx}
\usepackage{amsmath}
\usepackage{amsthm}
\usepackage{booktabs}
\usepackage{algorithm}
\usepackage{algorithmic}
\usepackage{tikz}
\usepackage{latexsym}
\usepackage{amssymb}
\usepackage{subcaption} 

\usepackage{natbib}
\def\cite#1{\citep{#1}}

\newcommand{\independent}{\mathrel{\text{\scalebox{1.2}{$\perp\mkern-10mu\perp$}}}}

\def\blue#1{{#1}}

\title{On Proximal Causal Learning with Many Hidden Confounders}

\author{
Nikos Vlassis$^1$\footnote{Contact Author}
\ \ \ 
Phil Hebda$^1$
\ \ \ 
Stephan McBride$^2$\footnote{The author was with Netflix when this work was concluded.}
\ \ \
Athanasios Noulas$^3$\footnotemark[2]\\
\affiliations
$^1$Netflix Research, CA, USA\\
$^2$Amazon.com, CA, USA\\
$^3$Traderepublic.com, Germany\\
\emails
\{nvlassis, phebda\}@netflix.com,
stemcb@amazon.com,
thanasis.noulas@traderepublic.com
}

\begin{document}

\maketitle

\begin{abstract}
We generalize the proximal g-formula of Miao, Geng, and Tchetgen Tchetgen (2018) for
causal inference under unobserved confounding using proxy variables.
Specifically, we show that the formula holds true for all causal models in a certain equivalence class, and this class contains models in which the total number of levels for the set of unobserved confounders can be arbitrarily larger than the number of levels of each proxy variable. Although straightforward to obtain, the result can be significant for applications. Simulations corroborate our formal arguments.
\end{abstract}

%%%%%%%%%%%%%%%%%%%%%%%%%%%%%%%%%
\section{Introduction}
%%%%%%%%%%%%%%%%%%%%%%%%%%%%%%%%%

\begin{figure*}[ht]
\begin{center}
  \begin{subfigure}[b]{0.2\textwidth}
    \centering
    \resizebox{\linewidth}{!}{
      \begin{tikzpicture}[
          observed/.style={circle, draw=black, fill=red!50, thick, minimum size=7mm},
          hidden/.style={circle, draw=black, fill=gray!50, thick, minimum size=7mm},
        ]
        \node[observed] (x) at (0,0) {$X$};
        \node[observed] (y) at (3,0) {$Y$};
        \node[hidden] (u) at (1.5,1.5) {$U$};

        \draw [->,>=latex] (x) edge (y);
        \path  [->,>=latex] (u) edge (x);
        \path  [->,>=latex] (u) edge (y);
      \end{tikzpicture}
    }
    \caption{~}
    \label{fig:subfig8}
  \end{subfigure}
  \qquad\quad
  \begin{subfigure}[b]{0.2\textwidth}
    \centering
    \resizebox{\linewidth}{!}{
      \begin{tikzpicture}[
          observed/.style={circle, draw=black, fill=red!50, thick, minimum size=7mm},
          hidden/.style={circle, draw=black, fill=gray!50, thick, minimum size=7mm},
        ]
        \node[observed] (x) at (0,0) {$X$};
        \node[observed] (y) at (3,0) {$Y$};
        \node[hidden] (u) at (1.5,1.5) {$U$};
        \node[observed] (z) at (0,1.5) {$Z$};
        \path  [->,>=latex] (z) edge (x);
        \draw [->,>=latex] (x) edge (y);
        \path  [->,>=latex] (u) edge (x);
        \path  [->,>=latex] (u) edge (y);
      \end{tikzpicture}
    }
    \caption{~}
    \label{fig:subfig8b}
  \end{subfigure}
  \qquad\quad
  \begin{subfigure}[b]{0.2\textwidth}
    \centering
    \resizebox{\linewidth}{!}{
      \begin{tikzpicture}[
          observed/.style={circle, draw=black, fill=red!50, thick, minimum size=7mm},
          hidden/.style={circle, draw=black, fill=gray!50, thick, minimum size=7mm},
        ]
        \node[observed] (x) at (0,0) {$X$};
        \node[observed] (y) at (3,0) {$Y$};
        \node[hidden] (u) at (1.5,1.5) {$U$};
        \node[observed] (w) at (3,1.5) {$Z$};

        \path  [->,>=latex] (x) edge (y);
        \path  [->,>=latex] (u) edge (x);
        \path  [->,>=latex] (u) edge (w);
        \path  [->,>=latex] (w) edge (y);

      \end{tikzpicture}
    }
    \caption{~}
    \label{fig:subfig9}
  \end{subfigure}
  \qquad\quad
  \begin{subfigure}[b]{0.2\textwidth}
    \centering
    \resizebox{\linewidth}{!}{
      \begin{tikzpicture}[
          observed/.style={circle, draw=black, fill=red!50, thick, minimum size=7mm},
          hidden/.style={circle, draw=black, fill=gray!50, thick, minimum size=7mm},
        ]
        \node[observed] (x) at (0,0) {$X$};
        \node[observed] (y) at (3,0) {$Y$};
        \node[hidden] (u) at (1.5,1.5) {$U$};
        \node[observed] (w) at (3,1.5) {$W$};
        \node[observed] (z) at (0,1.5) {$Z$};

        \path  [->,>=latex] (x) edge (y);
        \path  [->,>=latex] (u) edge (x);
        \path  [->,>=latex] (u) edge (y);
        \path  [->,>=latex] (u) edge (w);
        \path  [->,>=latex] (u) edge (z);
        \path  [->,>=latex] (w) edge (y);
        \path  [->,>=latex] (z) edge (x);

      \end{tikzpicture}
    }
    \caption{~}
    \label{fig:subfig10}
  \end{subfigure}
  \caption{Examples of causal graphs. The red nodes are observed
    variables, the grey node $U$ is unobserved. In (a) the variable $U$ acts as an unobserved
    confounder between $X$ and $Y$. In (b) the variable $Z$ acts as an
    instrument, adding exogenous variation to the system. In (c) the
    variable $Z$ can serve as a backdoor adjustment (i.e., admissible)
    variable to infer the causal effect of $X$ to $Y$
    (see Section \ref{methods}). In (d) the variables $Z$
    and $W$ can serve as proxies to infer the causal effect of
    $X$ to $Y$ (see Section \ref{negcontrols}).}
  \label{fig:subfig1.a.4}
\end{center}
\end{figure*}
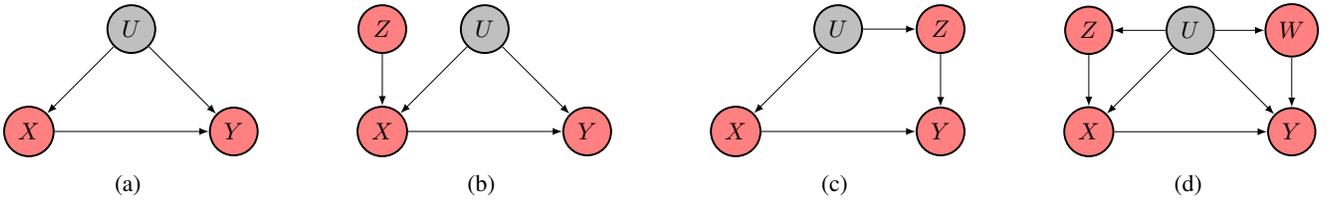

The gold standard in causal inference is randomized control trials. However, randomization can often be impractical, uneconomical, or even unethical, in which case the alternative is to 
infer the desired causal effects from observational data \cite{hernan2020causal,pearl2009causality,rubin2005causal}. In this paper we address the problem of causal inference from observational data when unobserved confounders are present. This is a hard problem in general, but a very relevant one for many applications.
In a seminal paper, Miao, Geng, and Tchetgen
Tchetgen \cite{miao2018identifying} demonstrated that the causal effect of a variable $X$ on some other variable $Y$ can be identified nonparametrically even under unobserved confounding, as long as a set of sufficient conditions are satisfied. Their approach, dubbed Proximal Causal Learning \cite{tchetgen2020introduction}, relies on the existence of two observed \emph{proxy variables} that must be associated to the latent confounders in a certain way. 
\blue{Proximal Causal Learning can be particularly useful for industrial applications, where domain knowledge is often leveraged for reasoning about possible hidden confounders, and a large number of potential proxies are available (e.g., from detailed information on users or items).}

\blue{As a motivating example from industry, consider users of an online service who perform a specific action, say interacting with a specific element in the service's UI. Let $X$ be a (binary) random variable that models the event that a user undertook the specific action, and let $Y$ denote the (binary) outcome that we are interested in studying, say if the user subsequently makes a purchase within the service.}
A causal quantity of interest is the \emph{average treatment effect}
\begin{equation}
\label{eq:ate}
P(Y=1  \mid  do(X=1)) - P(Y=1  \mid  do(X=0)) \, ,
\end{equation}
where the $do$ notation implies that we want to simulate the effect on $Y$ of \emph{fixing} $X$ to a specific value \cite{pearl2009causality}. The estimation of the effect in \eqref{eq:ate} can be difficult when unobserved confounders are present.
\blue{In our industry example, those confounders could be unobserved external sources of variation such as social buzz or other competing services, as well as unobserved user
characteristics such as demographics or price consciousness.}
Unless we account for such confounders, statistical inference of the treatment effect \eqref{eq:ate} may be prone to bias.

One way to account for an unobserved confounder $U$ is to employ two proxy variables, call them $Z$ and $W$, that are known to be coupled to $U$ in a certain way. In that case, and under certain (sufficient) conditions, the causal quantity $P(y \mid do(x))$ is \emph{nonparametrically identifiable} \cite{miao2018identifying,shi2020multiply,tchetgen2020introduction}. An example of a causal graph for which identification is possible is shown in Fig.~\ref{fig:subfig10}.

In the case where all variables are discrete, which we assume in this work, one of the conditions for identifiability in \citet{miao2018identifying} is that each of the two proxies $Z$ and $W$ must have the same number of levels (i.e., cardinality of their range) as the unobserved $U$. In this work we show that this condition can be significantly relaxed. As we elaborate in Section \ref{newresult}, the identification formula of \citet{miao2018identifying} holds true for all causal models in a certain equivalence class, and this class contains models in which the total number of levels for the set of unobserved confounders can be arbitrarily larger than the number of levels of $Z$ or $W$. This result, although easy to obtain, can be very important for applications: It opens the way to apply Proximal Causal Learning in a much wider class of problems than previously thought possible.

%%%%%%%%%%%%%%%%%%%%%%%%%%%%%%%%%%
\section{Causal graphs and $do$-calculus}
\label{methods}
%%%%%%%%%%%%%%%%%%%%%%%%%%%%%%%%%%

In this section we provide a short description of causal graphs and
$do$-calculus, mainly following \citet{bareinboim2016causal}. 
A causal graph is defined by a set $U$ of unobserved (latent) variables and a set $V$ of observed variables, 
which are assumed to be coupled by local causal dependencies
(deterministic functions) that give rise to a directed acyclic graph.  
For example, the causal graph of Fig.~\ref{fig:subfig8}
involves two observed variables $V = (X,Y)$ and a latent variable $U$,
which are coupled by local functions as indicated by the arrows.  

A causal graph allows to predict the effect on a variable~$Y$ of intervening on some other
variable $X$ by setting $X=x$.  In Pearl's
notation this is written as $P(y \mid do(x))$ and it corresponds to
the probability that $Y$ takes the value~$y$ in a modified causal graph in which
$X$ has been fixed to value~$x$ and all its incoming arrows in the
original graph have been removed.  
The quantity $P(y \mid do(x))$ can also be expressed as 
$P(y \mid do(x)) =
\sum_u Y_x(u) \ P(U=u)$, where $P(U)$ is the distribution of the
latent variable $U$, and $Y_x(u)$
is the potential outcome of unit $u$ had $u$ been
assigned treatment $X=x$
\cite{rubin2005causal}. 

A notable property of a causal graph is that,
regardless of the form of the coupling functions among variables and
regardless of the distribution $P(U)$ of the latent variables,
the distribution $P(V)$ of the
observed variables must obey certain conditional independence
relations, which can be characterized by means of a graphical
criterion known as \emph{d-separation}: 
A set $Z$ of nodes is said to block a path $p$ if either  
(i) $p$ contains at least one arrow-emitting node that is in $Z$ or 
(ii) $p$ contains at least one collision node that is outside $Z$ and
has no descendant in $Z$.
If a set $Z$ blocks all paths from a set $X$ to a set $Y$, then we say
that $Z$ \emph{d-separates} $X$
and $Y$, in which case it holds that $X$ and $Y$ are independent given $Z$,
which we write $X  \independent Y \mid Z$. For example, in Fig.~\ref{fig:subfig9} the variable $Z$
d-separates the variables $X$ and $Y$ in the induced graph in which
the outgoing arrow from $X$ is removed. 

Quantities such as $P(y \mid do(x))$ can be estimated
\emph{nonparametrically} from the observational distribution $P(V)$
using an algebra known as $do$-calculus  \cite{pearl2009causality}. The latter
is a set of rules that stipulate d-separation conditions in certain
induced subgraphs of the original graph. An example of $do$-calculus is
the `backdoor' criterion for specifying \emph{admissible} sets of
variables:
A set $Z$ is admissible for estimating the causal effect of $X$ on $Y$
if (i) no element of $Z$ is a descendant of $X$ and (ii) the elements
of $Z$ block all backdoor paths from $X$ to
$Y$ (those are paths that end with an arrow pointing to $X$). Using
counterfactuals, these two conditions imply $X  \independent Y_x \mid
Z$ (see \citet{bareinboim2016causal}), which is known as `conditional
ignorability' in the potential outcomes literature 
\cite{rosenbaum1983central}.
An admissible set allows expressing the interventional distribution
$P(y \mid do(x))$ via observational quantities that are directly
estimable from the data, using the g-formula \cite{hernan2020causal}:
\begin{equation}
\label{eq:backdoor}
     P(y \mid do(x)) = \sum_z P(y \mid x, Z=z) \ P(Z=z) \, .
\end{equation}
As an example, in Fig.~\ref{fig:subfig9} the variable $Z$ is
admissible because (i) it is not a descendant of $X$ and (ii) it
blocks the single backdoor path from $X$ to $Y$. Hence we can use
\eqref{eq:backdoor} to estimate
$P(y \mid do(x))$. Note that when all variables are discrete, the
required quantities in \eqref{eq:backdoor} amount to simple histogram
calculations.

The algebra of $do$-calculus has been shown to be complete, meaning that if the rules of $do$-calculus cannot
establish a way to write $P(y \mid do(x))$ as a functional of
$P(V)$, then the effect $P(y \mid do(x))$ is not identifiable
\cite{pearl2009causality}.

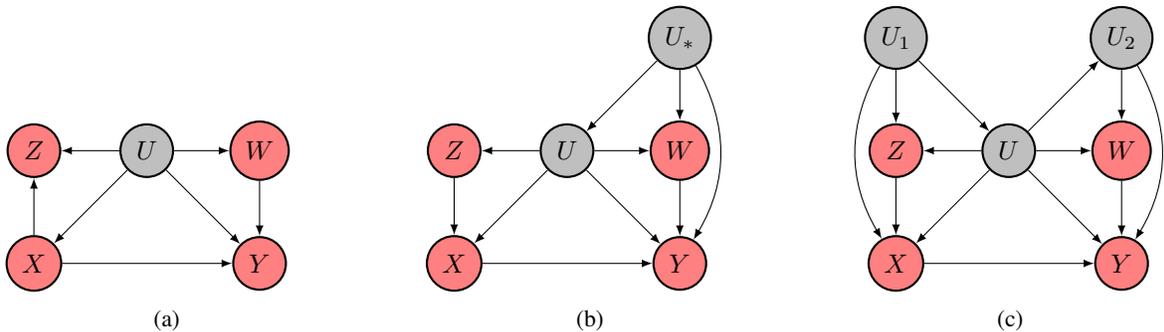
\begin{figure*}[t]
\begin{center}
    \begin{subfigure}[b]{0.25\textwidth}
           \begin{tikzpicture}[
observed/.style={circle, draw=black, fill=red!50, thick, minimum size=7mm},
hidden/.style={circle, draw=black, fill=gray!50, thick, minimum size=7mm},
]
    \node[observed] (x) at (0,0) {$X$};
    \node[observed] (y) at (3,0) {$Y$};
    \node[hidden] (u) at (1.5,1.5) {$U$};
    \node[observed] (w) at (3,1.5) {$W$};
    \node[observed] (z) at (0,1.5) {$Z$};

    \path  [->,>=latex] (x) edge (y);
    \path  [->,>=latex] (u) edge (x);
    \path  [->,>=latex] (u) edge (y);
    \path  [->,>=latex] (u) edge (w);
    \path  [->,>=latex] (u) edge (z);
    \path  [->,>=latex] (w) edge (y);
    \path  [->,>=latex] (x) edge (z);

\end{tikzpicture}
  \caption{~\label{fig:exotic1}}
\end{subfigure}
\qquad\quad 
    \begin{subfigure}[b]{0.25\textwidth}
\qquad 

\begin{tikzpicture}[
observed/.style={circle, draw=black, fill=red!50, thick, minimum size=7mm},
hidden/.style={circle, draw=black, fill=gray!50, thick, minimum size=7mm},
]
    
    \node[observed] (x) at (0,0) {$X$};
    \node[observed] (y) at (3,0) {$Y$};
    \node[hidden] (h) at (3,3) {$U_*$};
    \node[hidden] (u) at (1.5,1.5) {$U$};
    \node[observed] (w) at (3,1.5) {$W$};
    \node[observed] (z) at (0,1.5) {$Z$};

    \path  [->,>=latex](x) edge (y);
    \path  [->,>=latex](u) edge (x);
    \path  [->,>=latex](h) edge (u);
    \path  [->,>=latex](h) edge (w);
    \path  [->,>=latex](h) edge [bend left] (y);
    \path  [->,>=latex](u) edge (y);
    \path  [->,>=latex](u) edge (w);
    \path  [->,>=latex](u) edge (z);
    \path  [->,>=latex](w) edge (y);
    \path  [->,>=latex](z) edge (x);

\end{tikzpicture}
  \caption{~}
\label{fig:exotic2}
\end{subfigure}
\qquad\quad 
    \begin{subfigure}[b]{0.25\textwidth}
           \begin{tikzpicture}[
observed/.style={circle, draw=black, fill=red!50, thick, minimum size=7mm},
hidden/.style={circle, draw=black, fill=gray!50, thick, minimum size=7mm},
]

    \node[observed] (x) at (0,0) {$X$};
    \node[observed] (y) at (3,0) {$Y$};
    \node[hidden] (g) at (0,3) {$U_1$};
    \node[hidden] (h) at (3,3) {$U_2$};
    \node[hidden] (u) at (1.5,1.5) {$U$};
    \node[observed] (w) at (3,1.5) {$W$};
    \node[observed] (z) at (0,1.5) {$Z$};

    \path  [->,>=latex](x) edge (y);
    \path  [->,>=latex](g) edge (u);
    \path  [->,>=latex](u) edge (h);
    \path  [->,>=latex](h) edge (w);
    \path  [->,>=latex](h) edge [bend left] (y);
    \path  [->,>=latex](g) edge (z);
    \path  [->,>=latex](g) edge [bend right] (x);
    \path  [->,>=latex](w) edge (y);
    \path  [->,>=latex](z) edge (x);
    \path  [->,>=latex](u) edge (x);
    \path  [->,>=latex](u) edge (y);
    \path  [->,>=latex](u) edge (w);
    \path  [->,>=latex](u) edge (z);

\end{tikzpicture}
  \caption{~}
\label{fig:exotic3}
\end{subfigure}
\caption{Examples of causal graphs in the model equivalence class. The variables $U_*, U_1, U_2$ can have arbitrary many levels.
  See Section \ref{newresult} for details.} 
\label{fig:extendedgraph}
\end{center}
\end{figure*}

\section{Proximal Causal Learning}
\label{negcontrols}
%%%%%%%%%%%%%%%

In practical applications, it may be hard to identify an admissible
set of variables to use the backdoor g-formula
\eqref{eq:backdoor}, and reality may not be plausibly captured by a
graph such as the one in Fig.~\ref{fig:subfig9} (for example,
there might be omitted variables that  cause $U$ to directly affect
$Y$; that would appear in Fig.~\ref{fig:subfig9} as an extra
arrow from $U$ to $Y$). In such cases, identification of $P(y \mid
do(x))$ may nonetheless still be possible if a pair of \emph{proxy variables} (aka negative
  controls) are available, such as the variables $Z$ and
$W$ in the graph of Fig.~\ref{fig:subfig10}. Assuming all variables are discrete, the
corresponding identifiability result, due to Miao, Geng, and Tchetgen
Tchetgen \cite{miao2018identifying}, relies on the following assumptions:

\begin{description}
\item[1 Cardinalities -]
$|\mbox{Range}(Z)| = |\mbox{Range}(W)| = |\mbox{Range}(U)|$ (the variables $Z,W,U$ have equal number of levels).
\item[2 Backdoor -]
The unobserved variable $U$ is admissible, blocking all backdoor paths from $X$ to $Y$.
\item[3 Conditional independence -]
The observed variables $X,Y,Z,W$ and the unobserved $U$ jointly satisfy the following two conditional independence relations:
\begin{align*}
  \mbox{(i)} \quad & W  \independent (Z, X) \mid U \\
  \mbox{(ii)} \quad & Z \independent Y \mid (U, X) 
\end{align*}
\item[4 Matrix rank -]
The matrix $P(W \mid Z, x)$, whose $w$th row and $z$th column is $P(w \mid z, x)$, is invertible for each value $x$.
\end{description}
Note that condition 4 is testable from the data, whereas conditions 1--3 involve the unobserved variable $U$ and hence their validity must rely on domain knowledge about the area of study.
If all conditions 1--4 are met, then the quantity $P(y \mid do(x))$
is identifiable by the \emph{proximal g-formula}
\begin{equation}
\label{eq:miao}
     P(y \mid do(x)) = P(y \mid x,Z) \ 
    P(W \mid Z, x)^{-1} \,
    P(W) \, ,
\end{equation}
where $P(y \mid x,Z)$ and $P(W)$ are row and column vectors whose entries are $P(y \mid x,z)$ and $P(w)$, respectively, for all values of $z$ and~$w$. 
For a proof sketch of \eqref{eq:miao}, note that condition 2 allows us to write
\begin{equation}
 \hspace*{-5pt}  P(y \mid do(x)) 
    =  
       \sum_u P(y \mid u, x)  P(u) 
    =  
       P(y \mid U, x) P(U) 
\end{equation}
and the latter inner product can be re-expressed as
\begin{align}
  \hspace*{-5pt}       P(y \mid U, x)  
       \underbrace{P(U \mid x, Z) \,
       P(W \mid Z, x)^{-1} \,
       P(W \mid U)}_{\mbox{\small identity  matrix due to conditions 3,4}} 
       P(U) 
\end{align}
which, by condition 3(ii), simplifies to the RHS of \eqref{eq:miao}.

Note that the only difference of the proximal g-formula \eqref{eq:miao} with the g-formula
\eqref{eq:backdoor} is that the $P(Z)$ terms of \eqref{eq:backdoor} have been replaced with $P(W \mid Z, x)^{-1} \, P(W)$. 
In analogy with \eqref{eq:backdoor}, when all random variables are discrete,
the required quantities in \eqref{eq:miao} can be computed by simple histogram
calculations on the observed data. 
We refer to \citet{miao2018identifying,shi2020multiply,tchetgen2020introduction} for more
details and extensions.

\section{An equivalence class of models}
\label{newresult}
%%%%%%%%%

The full proof of \eqref{eq:miao} in \citet{miao2018identifying} reveals that  the condition 1 above (that $Z$, $W$ and $U$ have the same number of levels) is necessary for the proximal g-formula \eqref{eq:miao} to hold. This may at first seem too restrictive: The total number of levels for the unobserved $U$ may be hard to know in practice, and even if this number were somehow available, finding two proxy variables $Z$ and $W$ that simultaneously satisfy conditions 1, 3, and 4, can be a daunting task.   
Condition 1 can be relaxed to  $|\mbox{Range}(Z)| \geq |\mbox{Range}(U)|$,
and similarly for $W$, replacing the matrix inverse in \eqref{eq:miao} with a pseudoinverse \cite{shi2020multiply}. However, by increasing $|\mbox{Range}(Z)|$ and $|\mbox{Range}(W)|$  (e.g., by introducing more proxy variables) we may start violating conditions 3 and 4.
Fortunately, as we show next, condition 1 is not as restrictive as it may first appear.
 
Our key observation is that the proximal g-formula \eqref{eq:miao} must be true for all causal models in an equivalence class for which conditions 1--4 are true. This class contains more
expressive graphs than the graph shown in
Fig.~\ref{fig:subfig1.a.4}d. Critically, some of these graphs can involve additional latent nodes $U_*$ (see, e.g., Fig.~\ref{fig:extendedgraph}b), without violating conditions 1--4. 
This opens the avenue for applications in which $|\mbox{Range}(U_{\mbox{\tiny total}})|$, the total number of levels of the set of all latent confounders, is arbitrarily higher than $|\mbox{Range}(Z)|$ or $|\mbox{Range}(W)|$.
We do not attempt to provide a complete characterization of this equivalence class in this work; that would require different graphical tools \cite{jaber2019causal}.
Here we only slightly modify the set of conditions 1--4, and show examples of causal graphs that are significantly more expressive than the graph in Fig.~\ref{fig:subfig1.a.4}d, and for which the proximal g-formula \eqref{eq:miao} still holds. To maintain consistency with the notation we used in the previous sections, in the rest of the paper we will be using $U$ to denote a \emph{subset} of all latent confounders such that
$|\mbox{Range}(Z)| = |\mbox{Range}(W)| = |\mbox{Range}(U)|$, and we will be using different notation (e.g., $U_*$) for additional latent confounders.

Our modified set of conditions is obtained by re-expressing the backdoor criterion 2 by an
equivalent set of conditional independence relations as follows (see
\citet[Section 11.3.3]{pearl2009causality} for details). 
Let $T$ stand for the set of all
direct parents of $X$ (observed and unobserved), excluding
$U$ if $U$ is a direct parent of $X$ (in our generalization the node
$U$ need not be a direct parent of $X$). The set~$T$ may contain $Z$
and/or additional nodes not appearing in
Fig.~\ref{fig:subfig10}; see Fig.~\ref{fig:extendedgraph} for
examples. Then the backdoor criterion~2 can be replaced by the following:

\begin{description}
\item[2 Backdoor-surrogate]
\begin{align*}
  \mbox{(i)} \quad &  T \independent Y \mid (U, X) \\
  \mbox{(ii)} \quad  & X  \independent U \mid T \qquad \mbox{if $U$ is not a direct parent of $X$}
\end{align*}
\end{description}
Note that the above two conditions are subsumed by the single
conditional independence criterion 3(ii) when $T=Z$ and $U$ is a direct parent of $X$, as in
Fig.~\ref{fig:subfig10}. 

Given the above, the equivalence class is 
defined by the set of models that satisfy conditions 1 and 4 from Section~\ref{negcontrols}, together with the following set of graphical conditions (replacing conditions 2 and 3 from Section~\ref{negcontrols}):
\begin{description}
\item[Equivalence class:]
\begin{align*}
  \mbox{(i)} \quad & W  \independent (Z, X) \mid U \\
  \mbox{(ii)} \quad & Z \independent Y \mid (U, X) \\
  \mbox{(iii)} \quad & T \independent Y \mid (U, X) \\
  \mbox{(iv)} \quad  & X  \independent U \mid T \qquad \mbox{if $U$ is not a direct parent of $X$}
\end{align*}
\end{description}

In  Fig.~\ref{fig:extendedgraph} we show a few examples of graphs in the equivalence class that can capture  real-world dynamics and applications. The graph of Figure
\ref{fig:exotic1} uses a post-treatment variable $Z$ to perform
inference. Post treatment variables are known to bias regression
estimators, but they can be used in the Proximal Causal Learning framework as long
as the conditional independence relation $Z \independent Y \mid (U, X)$ holds \cite{shi2020multiply,tchetgen2020introduction}.

The graph of Fig.~\ref{fig:exotic2} has the potential to capture arbitrary high-dimensional confounding in $U_*$.
\blue{Returning to our industry example, this graph is particularly well suited for applications that involve interactions with each of the many elements within the UI: Engagement with each element can be studied independently by grouping the engagement with all other elements in $U_*$.}
That would be an example where we treat observed covariates as part of a latent $U_*$ in order to satisfy the sufficient conditions of the equivalence class for the proximal g-formula \eqref{eq:miao} to hold; alternatively we can condition on those covariates and use a modified estimator \citep{tchetgen2020introduction}. 

The graph of Fig.~\ref{fig:exotic3} allows for high-dimensional
$U_1$ and $U_2$, as long as they do not simultaneously influence the low-dimensional $U$ 
(i.e., $U$ is not a collider). 
\blue{In our running industry example, $U_2$ can capture engagement with UI elements other than the one being studied, while $U_1$ can capture competition, other UI elements, marketing, or word-of-mouth effects that influence how a user interacts with the service.}
In this model, $U$ can be viewed as a low-dimensional `bottleneck' from $U_1$ to $U_2$.

%%%%%%%%%%%%%%%%%%%%%%%%%%%%%%%%%%
\section{Simulations}
\label{simulation}
%%%%%%%%%%%%%%%%%%%%%%%%%%%%%%%%%%

Real-world observational settings lack both ground truth estimates of
the average treatment effect in~\eqref{eq:ate} and the ability to verify the
cardinality and conditional independence assumptions. We
can evaluate how well Proximal Causal Learning can recover causal effects through
simulated data with ground truth. 
Next we describe the structure of the simulations and how we leverage these simulations to understand properties
of a simple histogram-based estimator of \eqref{eq:miao} when the assumptions in Section~\ref{negcontrols} are violated.
Where relevant, we benchmark the results against a regression estimator.

\subsection{Simulation setup and notation}
\label{Section_A}
For grounding, we first discuss the simulation of the graphical model of Fig.~\ref{fig:subfig10}.
We will leverage and extend this simulation in the following subsections to handle the graphs in Fig.~\ref{fig:extendedgraph},
violations of conditional independence, and a high dimensional $U$.
All nodes are binary in all cases. We introduce
difference parameters $\delta_{ij}$ for the connection from node $i$ to node $j$,
e.g., $\delta_{UW} = P(W \mid U) - P(W \mid \neg U)$. From there on, given a
draw $u \sim U$ from a Bernoulli distribution with probability $\phi$, the values of the nodes are
populated according to:
\begin{align*}
& W \sim Ber(\phi + (u-\frac{1}{2}) \delta_{UW}) \\
& Z \sim Ber(\phi + (u-\frac{1}{2}) \delta_{UZ}) \\
& X \sim Ber(\phi + (u-\frac{1}{2}) \delta_{UX} + (z-\frac{1}{2}) \delta_{ZX}) \\
& Y \sim Ber(\phi + (u-\frac{1}{2}) \delta_{UY} + (w-\frac{1}{2}) \delta_{WY} +  (x-\frac{1}{2}) \delta_{XY} )
\end{align*}
where the average treatment effect is encoded by the difference parameter $\delta_{XY}$.

For additional details on the above structure, as well as extensions performed below, we refer to the
repository\footnote{The repository for the code that produced the simulations and their results is located at
\texttt{\url{https://github.com/hebda/proximal\_causal\_learning}}}.

\subsection{Bias in the model equivalence class}

\begin{table}[t]
\begin{center}
\begin{tabular}{|c|c|c|c|}
\hline
Graph & Regression & Proximal Causal Learning \\ \hline\hline
Fig.~\ref{fig:subfig10} & $24.4\% \pm 0.4\%$  & $\mathbf{-0.1\% \pm 1.7\%}$ \\ \hline %(26) \\ \hline 
Fig.~\ref{fig:exotic1} & $~~7.3\% \pm 0.3\%$  & $\mathbf{-0.0\% \pm 1.2\%}$ \\ \hline %(26) \\ \hline
Fig.~\ref{fig:exotic2} & $10.1\% \pm 0.4\%$ & $\mathbf{~~~0.1\% \pm 1.5\%}$  \\ \hline % (25) \\ \hline
Fig.~\ref{fig:exotic3} & $10.9\% \pm 0.5\%$ & $\mathbf{-0.2\% \pm 1.3\%}$  \\ \hline % (24) \\ \hline
\end{tabular}
\caption{\label{tab:exotic}Relative biases of estimators for the average treatment effect,
for simulations of the graphs of Fig.~\ref{fig:subfig10} and Fig.~\ref{fig:extendedgraph}
using regression and Proximal Causal Learning. Regression leverages all observed nodes in the given graph.
Bold font indicates the best performance. We show empirical mean and variance from 100 runs of the simulated graph.}
\end{center}
\end{table}

Table~\ref{tab:exotic} contains the results for the graphs of Fig.~\ref{fig:extendedgraph}, including the base case of Fig.~\ref{fig:subfig10}.
We extend the simulation to include additional nodes $U_*$, $U_1$, and $U_2$ where applicable.
We compare the Proximal Causal Learning method against a baseline regression
approach that uses the same covariates. The comparison is between each
estimator's relative bias, which is defined as the relative difference between
the estimated and true average treatment effect in~\eqref{eq:ate}.
Proximal Causal Learning gives an effectively unbiased
result for the three graphs in Fig.~\ref{fig:extendedgraph}, with the small, non-zero values coming
from sample variation. The regression benchmark is significantly biased in all cases; however we note that
regression is typically able to leverage many more covariates, which may result in a reduced, though non-zero, bias,
depending on the application.

\subsection{Implications of the invertibility of $P(W \mid Z,x)$}
\label{invertible}
The invertibility of $P(W \mid Z,x)$ (criterion~4 from Section~\ref{negcontrols}) can be directly tested from the data by
looking at the condition number of each $P(W \mid Z,x)$. For the 2-dimensional
case, we have two matrices to consider, one for each value of $x$, and
we use the larger of the two condition numbers to describe the
stability of the matrix inversion.

\begin{figure}[h]
\includegraphics[width=\linewidth]{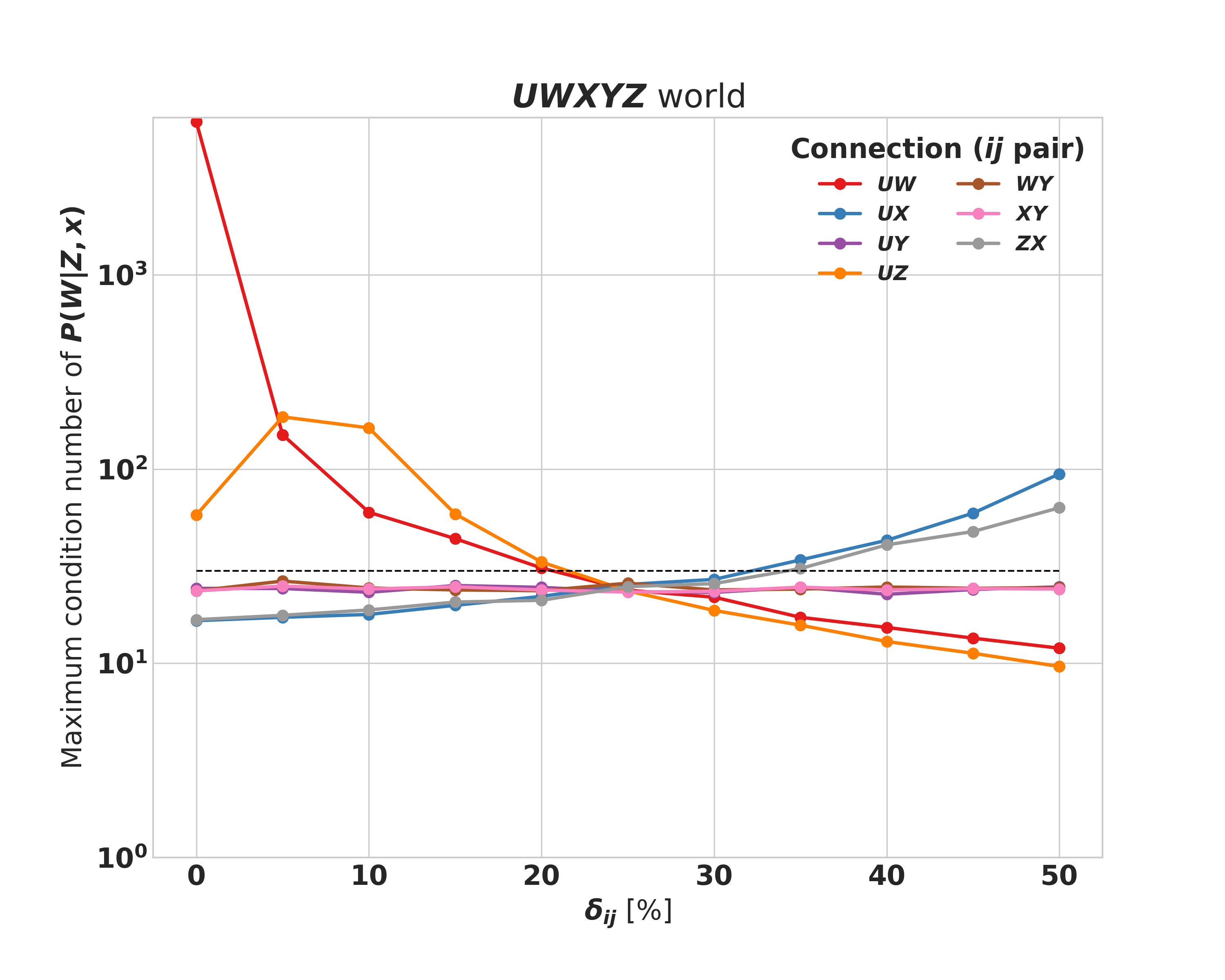}
\caption{Dependence of the condition number of $P(W \mid Z,x)$ on the
  underlying conditional probabilities. The dashed line represents a
  reference cut-off of 30.\label{fig_condition_number}} 
\end{figure}

In Fig.~\ref{fig_condition_number} we can see that requiring a
condition number of 30, which would lead to stable matrix inversion, corresponds to an implicit claim on the strength of
certain connections, captured through the $\delta$s in the structure
of the simulation. These are that $U \rightarrow W, Z$ cannot be
arbitrarily small, and that $U,Z \rightarrow X$ cannot be arbitrarily
large. Furthermore, the connections $U,W,X \rightarrow Y$ can be of
arbitrary strength. In short, the unobserved confounder must be
sufficiently captured by the proxies, without exerting too much force
on the treatment condition.

The requirement of having a maximum condition number of 30 is somewhat arbitrary and context dependent.
Ultimately, this quantity is related to the variance of the estimator. Numerically stable matrix inversion provides
low variance, while unstable inversion provides the opposite.

\subsection{Violations of the independence assumptions}
\label{indep}
In practical applications, the selection of variables $W$ and $Z$ is
very important since the independence assumptions are not testable. In temporal data, one could use certain future
observations as proxies, as the proxies do not need to be pre-treatment \cite{shi2020multiply}.
\blue{In industry settings, variables $Z$ are generally easier to discover: Most interventions can only affect an outcome of interest through user interactions with the service. Variables $W$ are generally harder to find, as they should affect the outcome without affecting interactions with the service.}

In this subsection, we examine the
bias in the average treatment effect that is induced into an estimator of~\eqref{eq:miao} when the
independence assumption $W  \independent (Z,X) \mid U$ is violated, and we
benchmark this bias against a regression estimator. We impose this
violation into the simulation by modifying the $X$-step in the
simulation to include the term $(w - \frac{1}{2}) \delta_{WX}$.
In this way, $X$ is conditionally dependent on $W$ in addition to $U$ and $Z$.

\begin{figure}[h]
\includegraphics[width=\linewidth]{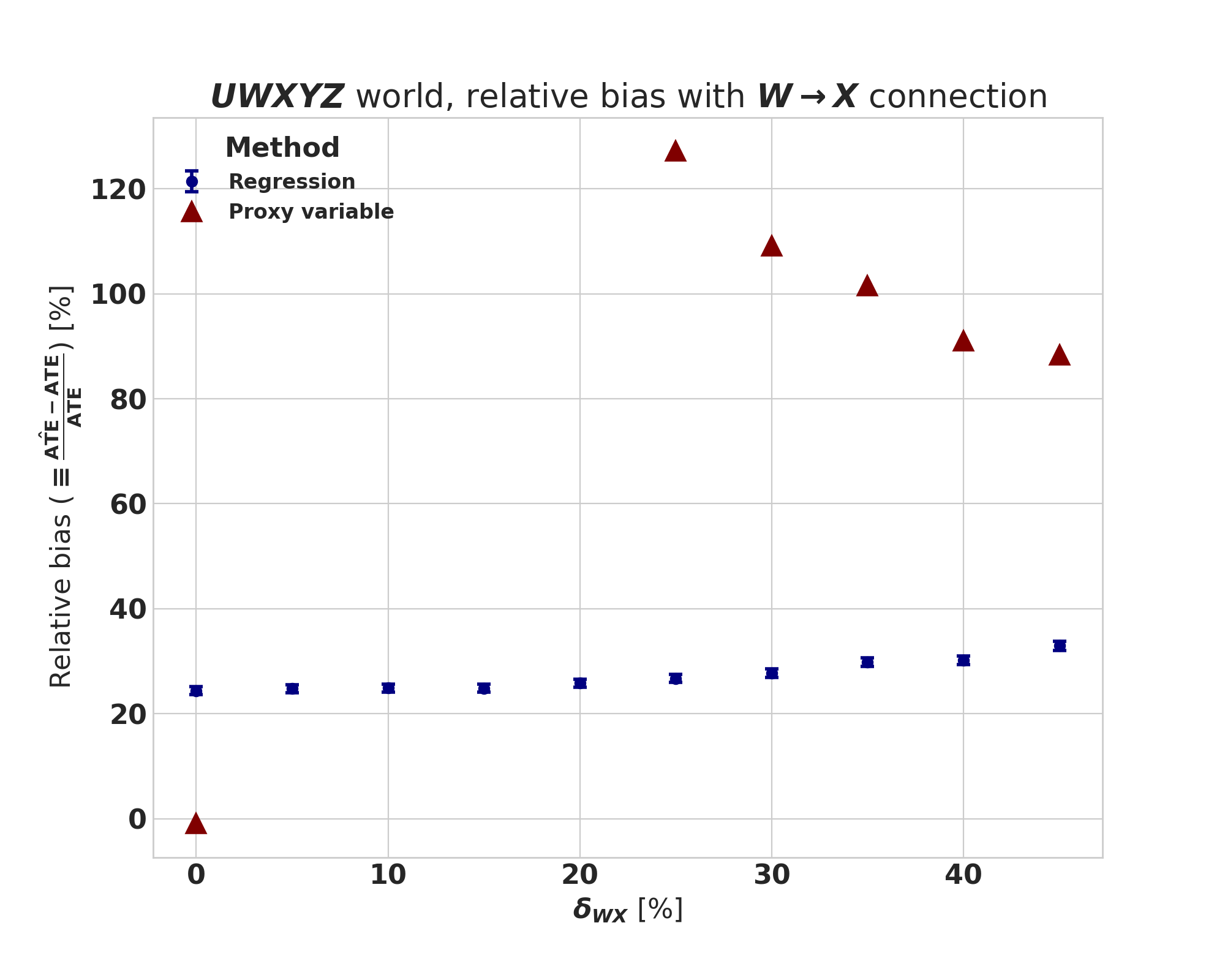}
\caption{\label{fig_bias} Bias in the estimation of the average treatment effect as a function 
$\delta_{WX}$. The regression estimator is provided with 95\% confidence intervals.
The proximal g-formula estimate \eqref{eq:miao} is
shown only when the condition number of $P(W \mid Z,x)$ does not
exceed 30; otherwise the estimate is omitted.}
\end{figure}

Fig.~\ref{fig_bias} shows the bias of the two estimators for
arbitrary $\delta_{WX}$. We see that the bias for regression is
roughly independent, though non-zero, for all values of $\delta_{WX}$. For
Proximal Causal Learning, we see that the bias is strongly dependent on the
strength of the $W \rightarrow X$ connection. Careful selection of the
$W$ variable will lead to a variable for which $\delta_{WX}$ is not
large. Additionally, the invertibility of $P(W \mid Z,x)$ provides
protection against arbitrarily high values of bias, since large values
of $\delta_{WX}$ cause this testable criterion to fail. (Such points are omitted from the plot.)

\subsection{Bias when the dimensionality of $U$ is large}
\label{dimensionality}

\begin{table*}[t]
\begin{center}
\begin{tabular}{|l|c|c|}
\hline
Simulation Setup & Regression & Proximal Causal Learning\\ \hline\hline
Fig.~\ref{fig:exotic2}, binary $X$ & $10.0\% \pm 1.9\%$ &
                                                            $\mathbf{~~~0.7\%}$  \\ \hline
Fig.~\ref{fig:exotic2}, 5-dimensional $X$ & $10.5\% \pm 9.9\%$ &
                                                            $\mathbf{~~~1.0\%}$  \\ \hline  
Fig.~\ref{fig:subfig10}, $\delta_{UV,j}$ constant in $j$ & $42.8\% \pm 4.1\%$ & $\mathbf{-2.0\%}$ \\ \hline % (11) \\ \hline
Fig.~\ref{fig:subfig10}, $\delta_{UV,j}$ decreases linearly in $j$  & $18.7\% \pm 4.0\%$ & $\mathbf{-1.9\%}$ \\ \hline % (28) \\ \hline
Fig.~\ref{fig:subfig10}, $\delta_{UV,j}$ decreases linearly in $j$, $\delta_{UY,j}$ constant in $j$ & $18.4\% \pm 4.1\%$ & $\mathbf{-0.2\%}$ \\ \hline % (28) \\ \hline
Fig.~\ref{fig:subfig10}, $\delta_{UV,j}$ constant in $j$, $\delta_{UY,j}$ decreases linearly in $j$ & $42.9\% \pm 4.0\%$ & $\mathbf{-1.3\%}$ \\ \hline % (11) \\ \hline
Fig.~\ref{fig:subfig10}, $\delta_{UV,j} = 0$ for $j \ge 1$, $\delta_{UY,j}$ constant in $j$ & $~~4.6\% \pm 3.9\%$ & $\mathbf{-0.3\%}$ \\ \hline % (27) \\ \hline
Fig.~\ref{fig:subfig10}, $\delta_{UV,j}$ constant in $j$, $\delta_{UY,j} = 0$ for $j \ge 1$ & $40.4\% \pm 3.8\%$ & $\mathbf{~~~2.4\%}$ \\ \hline % (10) \\ \hline
\end{tabular}
\caption{\label{tab_high_dim_U} Relative bias of the average treatment effect
for graphs with high dimensional $U$ (10 dimensions, a total of 1024 levels). Bold fonts indicate
  best performance. We present 95\% confidence intervals
with the regression result; at present, there is no analytical means to estimate
the variance of the Proximal Causal Learning estimator.}
\end{center}
\end{table*}

In most applications, it is hard to have a good estimate of the
dimensionality of the unobserved confounder $U$. In addition, the variables $W$ and $Z$
need to have the same number of levels as the total number of levels of $U$, for the matrix $P(W \mid Z,x)$ to be
invertible.
In this subsection, we examine the bias under a high
dimensional $U$ that violates the (untestable) cardinality
assumption~1 for the proximal g-formula \eqref{eq:miao}. We focus on the graphs of Figures~\ref{fig:subfig10} and \ref{fig:exotic2}.

In this simulation we maintain binary $W$ and $Z$ and allow $U$ to
contain a total of 10 binary dimensions. This produces an
unobserved confounder space with a total of 1024 levels. We encode difference parameters $\delta_{ij}$
for each binary dimension by treating the relevant variables as 10-dimensional vectors rather than scalars.
Table~\ref{tab_high_dim_U} presents the results under different types of confounding effects. 
When comparing with Table~\ref{tab:exotic} we see that, when the 
dimensionality of $U$ increases to 10, our estimates increase in bias, and this bias
does not depend on the dimensionality of $X$. Nevertheless, in all types of
confounding we simulated, this bias is
relatively small for Proximal Causal Learning, and it is consistently smaller than when using regression.

\section{Conclusions and discussion}
\label{nextsteps}

We have shown that the identification result of Miao, Geng, and Tchetgen
Tchetgen \cite{miao2018identifying}
for Proximal Causal Learning under unobserved confounding holds true for all causal models in a certain equivalence class. %
This class contains models in which the total number of levels for the set of unobserved confounders can be arbitrarily higher than the number of levels of the two proxy variables, an important result for industry applications.

In the simulations we have also studied a number of different properties of a simple histogram-based estimator of \eqref{eq:miao}, such as its sensitivity to misspecification or violation of some of the sufficient conditions. The simulations have additionally identified that, for low
dimensional proxies, even the presence of a high dimensional $U$ will yield less biased estimates under the Proximal Causal Learning estimator than with standard regression methods.
Overall, our results provide further evidence that Proximal Causal Learning can be a very promising causal inference method in an observational setting.  

An interesting open question is to completely characterize the equivalence class of models for which identification holds (using tools such as ancestral graphs \cite{jaber2019causal}), and to establish a necessary condition for identifiability. Another useful direction
is to derive the population moments of an estimator of \eqref{eq:miao}, both under a specified
as well as under a misspecified causal graph model, and see how the risk of the estimator varies with
the problem inputs (for instance, what is the precise dependence of the variance of the estimator 
on properties of $P(W \mid Z, x)$, as we have alluded to above).
Our simulations are already providing some answers to those questions. Nonetheless, 
an analytical treatment would be valuable as it would offer more intuition into the
applicability of Proximal Causal Learning in practical causal inference problems, especially those often encountered
in the industry. 

\bibliographystyle{apalike}
\bibliography{ms}

\end{document}